\documentclass[iop,apj,twocolumn]{emulateapj}
\usepackage{apjfonts}
\usepackage{natbib}
\usepackage{amsfonts}
\usepackage{amsmath}
\usepackage{multirow}
\usepackage{enumerate}
\usepackage[normalem]{ulem}
\usepackage{graphicx}

\usepackage{epsfig}
\usepackage{rotating}
\usepackage{float}
\usepackage{verbatim}

\usepackage{wrapfig}

\newcommand{\omg}{\Omega}
\newcommand{\beq}{\begin{equation}}
\newcommand{\eeq}{\end{equation}}
\newcommand{\bea}{\begin{eqnarray}}
\newcommand{\eea}{\end{eqnarray}}

\newcommand{\eps}{\epsilon}
\newcommand{\rrat}{r_\textrm{ratio}}
\newcommand{\Atil}{\tilde{A}}
\newcommand{\Atc}{\tilde{A}_\textrm{crit}}
\newcommand{\Btil}{\tilde{\beta}}
\newcommand{\Hm}{H_\textrm{max}}
\newcommand{\Em}{\eps_\textrm{max}}

\begin{document}

\title{A new view on the maximum mass of differentially rotating neutron stars}
\author{D. Gondek-Rosi\'nska\altaffilmark{1,*}, I. Kowalska\altaffilmark{2}, L. Villain\altaffilmark{3}, M. Ansorg\altaffilmark{4}, M. Kucaba\altaffilmark{1}}
\altaffiltext{1}{Janusz Gil Institute of Astronomy, University of Zielona G\'ora, Licealna 9, 65-407, Zielona G\'ora, Poland} 
\altaffiltext{2}{Astronomical Observatory Warsaw University, 00-478 Warsaw, Poland}
\altaffiltext{3}{Laboratoire de Math\'ematiques et Physique Th\'eorique, Univ. F.~Rabelais - CNRS (UMR 7350), F\'ed. D. Poisson, 37200 Tours, France}
\altaffiltext{4}{Theoretisch-Physikalisches Institut, Friedrich-Schiller-Universit\"at Jena, Max-Wien-Platz 1, D-07743 Jena, Germany}
\altaffiltext{*}{E-mail: dorota@astro.ia.uz.zgora.pl}

\date{Accepted XXX. Received XXX; in original form XXX}

\begin{abstract}
We study the main astrophysical properties of differentially rotating neutron stars described as stationary and axisymmetric configurations of a moderately stiff $\Gamma=2$ polytropic fluid. The high level of accuracy and of stability of our relativistic multidomain pseudo-spectral code enables us to explore the whole solution space for broad ranges of the degree of differential rotation, but also of the stellar density and oblateness. Staying within an astrophysicaly motivated range of rotation profiles, we investigate the characteristics of neutron stars with maximal mass for all types of families of differentially rotating relativistic objects identified in a previous article \citep{AGV}. We find that the maximum mass depends on both the degree of differential rotation and on the type of solution. It turns out that the maximum allowed mass can be up to 4 times higher than what it is for non-rotating stars with the same equation of state. Such values are obtained for a modest degree of differential rotation but for one of the newly discovered type of solutions. Since such configurations of stars are not that extreme, this result may have important consequences for the gravitational wave signal to expect from coalescing neutron star binaries or from some supernovae events.
\end{abstract}


\section{Introduction}

Binary neutron star (BNS) mergers are thought to be promising sources of gravitational waves and of neutrinos \citep{R15, Bernuzzi16}, as well as the progenitors of short gamma ray bursts \citep*{BNPP84,ELPS89}. After the detection of gravitational waves from binary black holes by the LIGO experiment \citep{Abbotta, Abbottb}, they are even one of the most expected next targets. The outcome of a BNS merger is the formation of a stellar black hole, but the latter can either occur promptly or be slightly delayed and include a stage during which a massive and warm differentially rotating neutron star is produced \citep{SU02}. Whatever is the actual scenario, highly sophisticated and realistic numerical simulations are needed to ascertain the signals to be expected and consequently to enable us to extract information on both the gravitational and the high energy underlying physics. Although a lot of progress has been made, building numerical codes that take into account all the pertinent physical ingredients is such a difficult task that it is still far from achieved, making useful the resort to simplified studies concentrating on some specific aspects. One basic but key issue is the maximum life span of the potential short-lived material remnant, a question which can be approached by focussing first on its maximum mass. The analysis of the involved timescales and of the results of numerical simulations \citep{STU05} shows that a rough but reasonable approximation to address this problem consists in modeling the central body as a stationary axisymmetric rotating relativistic star in differential rotation, neglecting complicated inner motions of the matter, nuclear reactions, thermal effects, magnetic fields, etc. Doing so, it is for instance possible to study the influence of the degree of differential rotation or of the stiffness of the equation of state (EOS) on the maximum mass which, for rotating stars, can be much higher than for static stars [see for instance \cite{CST92,CST94a,CST94b,BSS0,LBS03}].

In a previous article \citep{AGV} (later referred to as Paper~I), a new investigation was presented of the structure of differentially rotating neutron stars, modelized as constant density stars or relativistic $N=1\,(\Gamma=2)$ polytrops. This study, extended for other polytropic EOSs in \cite{SKGVA}, relied on a multi-domain spectral code \citep*[based on the so called ``AKM-method'',][]{AKM} that was formerly shown to enable the calculation of very extremal configurations of rigidly rotating relativistic stars \citep*{AKMb,SA} or rings \citep*{Ansorg2005,AP05}. In Paper~I, only star-like configurations, \emph{i.e.} with a spheroidal topology (without a hole), were considered, but allowing what are sometimes called "quasi-toroidal" configurations in which the maximal density is not the central one. The focus was put on the solution space, and a noticeable result was the discovery of four "types" of configurations that co-exist with each other even for reasonable profiles of angular momentum.

The purpose of the present article is to extend the study of Paper I by calculating, for $\Gamma=2$ polytrops, astrophysicaly relevant quantities, such as the maximal mass, the angular momentum, the ratio between kinetic and potential energies, etc. Our highly accurate and stable spectral code enable us, for the first time, to study in detail those properties of differentially rotating neutron stars taking into account the whole solution space identified in Paper~I. Moreover, the understanding of the global structure of the parameter space makes it possible to explain the results of preceding studies, especially the works by \cite{BSS0} and \cite{LBS03}, showing that strange features of some sequences they had obtained arise from the fact that their codes were jumping from one type of configuration to another due to numerical limitations. Finally, the configurations we have calculated could be used as initial data to perform in a systematic way the stability analysis of differentially rotating neutron stars and to determine stability criteria for such objects.

The plan of the article is as follows: in Section~\ref{s:rotlaw}, we start by recapitulating the issue of differential rotation in relativistic stationary rotating stars, with as a primary goal to describe the hypothesis done in our work and to define variables and notations. Then, in Section~\ref{s:mass}, we briefly review current knowledges concerning the maximal mass of rotating relativistic stars and we present our results in the context of the existence of the various types that were introduced in Paper I. Later on, Section~\ref{s:rotpr} focusses on angular momentum and other quantities related to rotation and stability after which Section~\ref{s:conc} summarizes the results achieved, in contrast with those of previous studies. Finally, Appendix~\ref{s:NumScheme} reviews the feature of the numerical code, displaying tests of convergence and accuracy of our results, putting emphasis on the specificities of the version used in this article as well as in Paper~I.


\section{Models of differentially rotating relativistic stars}\label{s:rotlaw}

Stationary and axisymmetric configurations of rotating relativistic
stars have already been the subject of numerous works, be they
analytical or numerical, making this topic quite classical
\citep[see][for reviews]{S03,FS}. Since in this article we use exactly
the same equations and notations as in Paper~I, we send the reader
to this article for more detail. Here, we shall only remind the main assumptions concerning matter and its motion, in the framework of differentially rotating relativistic stars. 

As in Paper I, we work with units such as \mbox{$c=G=K=1$}, where $c$ is the
speed of light in vacuum, $G$ is Newton's constant and $K$ is the polytropic constant. The latter is defined by the polytropic equation of state, $p\,=\,K\,\eps_B^\Gamma\,$, with $\Gamma=2$ in this article, which relates the pressure $p$ to the rest-energy (or baryonic energy) density $\eps_B$. For reasons that will be reminded further, the main thermodynamical variable that shall be used is $H$, the dimensionless relativistic enthalpy defined from the pressure $p$ and the total energy density $\epsilon$ as
\beq\label{e:defH}
H(p)\,=\,\int_0^p\,\frac{dp}{\epsilon(p)+p}\,.
\eeq
To smooth the way to comparison with other codes and systems of units, we remind the reader
that for a \mbox{$\Gamma\,=\,2$} polytropic equation of state with
null temperature, the enthalpy~(\ref{e:defH}) verifies, in units
$G=c=1$ (but without any fixed values of $K$ for the time being),
\beq
H\,=\,\log\left(1\,+\,2\,K\,\eps_B\right)\,,
\eeq
while the total energy density $\epsilon$, that will be used in the following, satisfies
\beq
\epsilon(\epsilon_B)\,=\,\epsilon_B\,+\,K\,\epsilon_B^2\,,
\eeq
so that
\beq
\epsilon(H)\,=\,\frac{1}{4K}\left(e^{2H}\,-\,1\right)\,.
\eeq

Describing the rotational properties is made much
easier by introducing the fluid angular velocity with respect to
infinity, $\omg\,=\,u^{\phi}/u^t$, where the $u$-variables are the
non-zero components of the four-velocity, and an auxiliary variable
that we shall write $Y$, defined as $Y\,=\,u^t\,u_{\phi}$ and
which is some kind of specific angular momentum of the fluid. While
the rotation profile of stationary and axisymmetric rotating Newtonian
barotrops has to be such that the angular velocity only depends on the
distance from the axis of rotation \citep{G73}, it is a well-known fact
\citep{B70,BI76}, that for a relativistic one, conservation of the
energy-momentum tensor implies the condition $Y\,=\,F(\omg)\,,$ where
$F$ is an arbitrary function. Such a weak constraint allows for
numerous possibilities, some of which were for instance explored in
\citet{GYE12,UTGHS16}. Here we should however restrict ourselves to the now
classical law proposed by \citet*{KEH}, already used by many authors
\citep[among which][]{CST92,BGSM,GHZ98,BSS0,LBS03,MBS04,VPCG04} and in Paper~I:
\beq \label{e:lawdif1}
F(\omg)\,=\,A^2\,(\omg_c\,-\,\omg)\,,
\eeq
where $A$ is a parameter with the dimension of a length, while it can be shown than $\omg_c$ is the limit of $\omg$ on the rotation
axis. More precisely, we shall put Eq.~\eqref{e:lawdif1} in the form
\beq \label{e:lawdif2}
\omg\,=\,A_\omg(Y)\,=\,\omg_c\,-\,\frac{Y}{(\hat{A}\,r_e)^2}\,,
\eeq
where we introduced the equatorial radius of the star $r_e$, which is a
typical lengthscale of the problem, and \mbox{$\hat{A}=A/r_\mathrm{e}$}. Notice that since the rotation profile
tends to be rigid (or uniform) when $A \to \infty$ ($r_e$ keeping a finished
value), we follow \cite{BSS0} and parametrize the sequences and the degree of
differential rotation with
\beq\label{e:tilde_A}
\Atil=\hat{A}^{-1}=r_\mathrm{e}/A\,.
\eeq
We send the reader to Paper I and references therein for more details about the basic (astro)physical consequences of this law (Newtonian limit, etc.).

For a stationary and axisymmetric barotropic perfect fluid in rotation, Bianchi identities (or equivalently the relativistic Euler's equation) imply that there is a first integral of motion, that we choose to write
\beq \label{e:euler}
H\,=\,V_0\,-\,V\,-\,B_\omg(Y)\,,
\eeq
where:
\begin{itemize}
\item[i)] $H$ is the relativistic enthalpy~\eqref{e:defH};
\item[ii)] $V$ is defined by
\beq \label{e:defV}
u^t\,=\,e^{-V}\,,
\eeq
$V_0$ being its value at the North pole;
\item[iii)] and
\beq
B_\omg(Y)\,=\,\int_0^Y\,x\,\frac{d A_\omg(x)}{dx}\,dx\,,
\eeq
with $A_\omg$ introduced in Eq~(\ref{e:lawdif2}).
\end{itemize}

Working with the law~(\ref{e:lawdif1}), a configuration of a star corresponds to a solution of the first integral of motion together with the Einstein equations and a given equation of state ($\Gamma=2$ polytrop here), which is obtained by fixing three parameters such as the maximal enthalpy $H_m$, the central angular velocity $\Omega_c$ and the $\Atil$ quantity. Due to the non-linear nature of this system of equations, there is nevertheless not always uniqueness of the solution if those are indeed the fixed quantities, a feature that makes the solution space even richer than it is for rigidly rotating stars, as was shown and discussed in Paper~I. Furthermore, we remind that a sufficiently high degree of differential rotation can make the enthalpy (or equivalently some density) taking its maximal value outside of the symmetry axis, so that its central value is not an optimum parameter to fix. In the code used for this work, the resort to a Newton-Raphson scheme enables us to easily change the fixed parameters (see the Appendix~\ref{s:NumScheme}) and to explore the solution space by freely varying any parameters. Among the convenient quantities adapted to uniquely specify a fast rotating star are the ratio between its polar and equatorial radii, $\rrat=r_p/r_e$, with $0<\rrat<1$, and the rescaled shedding parameter $\Btil$ defined in Paper~I by
\beq\label{eq:betat}
\Btil=\frac{\beta}{\beta+1}\,,
\eeq
with \citep[see][]{AKMb}
\beq\label{eq:beta}
\beta=-\frac{r_e^2}{r_p^2}\frac{d(z_{b}^2)}{d(\rho ^2)}\mid_{\rho=r_e}\,,
\eeq
where the equation $z=z_b(\rho)$ describes the surface of the star. It can be verified that $0<\Btil<1$, that $\Btil \to 1$ when a star enters into the toroidal regime ({\it i.e.} $\rrat \to 0$), that $\Btil \to 0$ in the mass-shedding limit and that $\Btil = 1/2$ for a non-rotating spherical star.

As stated in the Introduction, we shall focus in this article on the issue of the structure of differentially rotating relativistic stars with maximal mass, which will be the topic of the next Section. Having in mind astrophysical scenarios such as the collapses or the mergers described earlier, we shall in the following also briefly discuss quantities related to rotation and instabilities, such as the angular momentum, the ratio between kinetic and gravitational energies, or the Kerr parameter.


\section{Maximum mass of neutron stars}\label{s:mass}

For a given equation of state, the maximum mass of non-rotating neutron stars ($M_{\rm max,\,stat}$) is obtained by solving the Tolman-Oppenheimer-Volkoff (TOV) equations, varying the value of the central energy density $\eps_{\rm c}$. Calculations show that, for realistic EOSs, it belongs to the range 2-2.5 $M_\odot$. In the case of rigid rotation, the centrifugal force enables its value to be higher by 12\%-20\% for neutron stars \citep[e.g.][]{CST94a,CST94b} and by $\sim 35$\% for strange stars \citep{G0,G99}. Other studies established that differential rotation can even be much more efficient, making possible an increase of the maximum mass by more than 60\%, especially for moderately stiff polytropic EOSs \citep{BSS0,LBS03}. As we will explain, our work demonstrates that, in the same conditions as considered in those previous studies, the actual effect of differential rotation can in fact be stronger.

\subsection{Comparison with previous studies}

\begin{figure*}
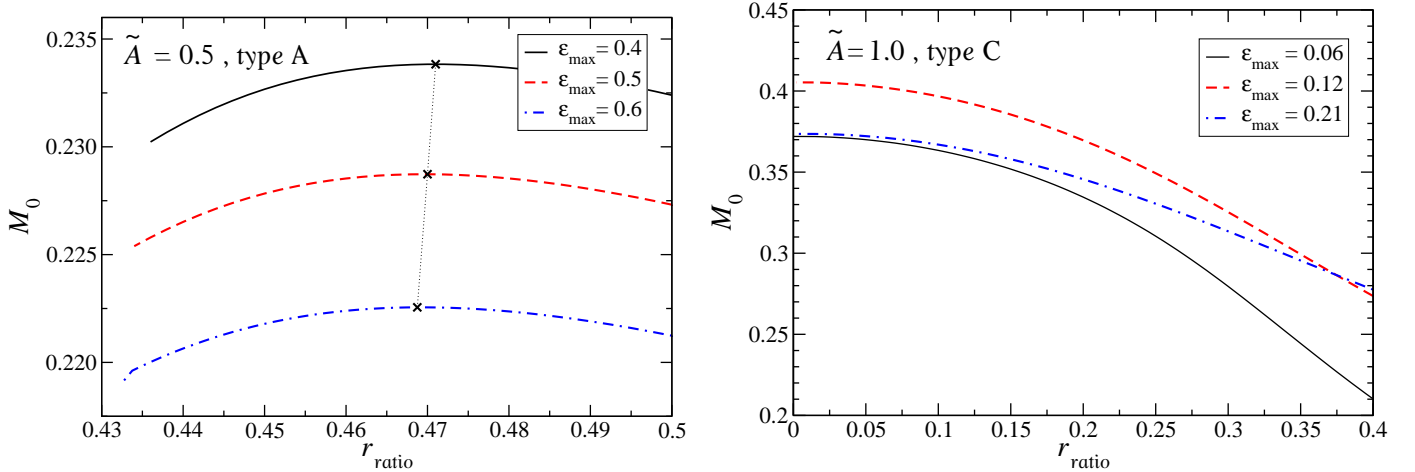


\begin{minipage}[b]{.5\textwidth}
\begin{centering}
\includegraphics[width=\textwidth,angle=0]{A05_M0_ratio}
\end{centering}
\end{minipage}
\hspace{0.15cm}
\begin{minipage}[b]{.5\textwidth}
\begin{centering}
\includegraphics[width=\textwidth,angle=0]{A1_M0_ratio.eps}
\end{centering}
\end{minipage}
\caption{Rest mass versus the ratio of the polar to equatorial radius along sequences with fixed energy density $\Em$ and close to the configuration with the maximum mass in the sequence. The left panel corresponds to sequences of configurations with a modest degree of differential rotation ($\Atil=0.5$), classified as type A configurations, while the right panel displays sequences for $\Atil=1.0$, which are of type C. For each sequence of type A, the configuration with the maximum mass is marked with a {\bf cross} and the terminal configuration (with the smallest value of $\rrat$) is at the mass-shedding limit. For type C sequences, we arbitrarily end the sequence when $\rrat=0$ (hence considering only stars with a spheroidal topology but with a toroidal shape, see Fig.~\ref{shapeAC}) and the maximum mass was found to be reached for such configurations.\label{typeAC}}

\end{figure*}

\begin{figure}
\begin{center}
\includegraphics[width=0.5\textwidth,clip,angle=0]{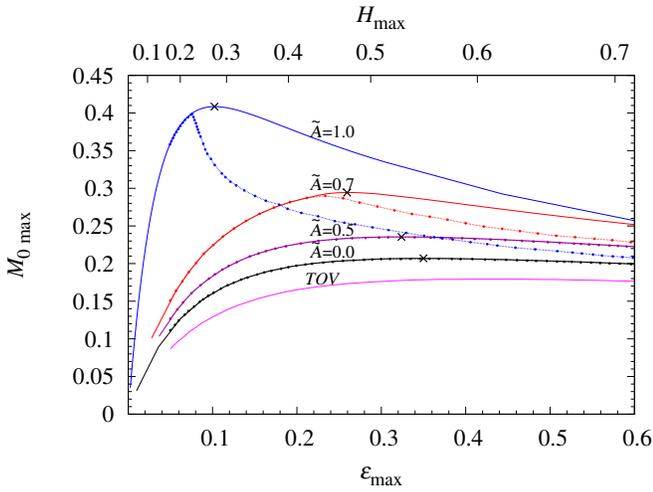}
\caption{Upper limits on the maximum rest mass $M_0$ versus the maximum energy density ($\Em$) (or maximum enthalpy $\Hm$) for differentially rotating neutron stars described by the $\Gamma=2$ polytropic equation of state for three fixed values of $\Atil$ ($0.5$, $0.7$ and $1.0$). For comparison, the results for static stars (TOV) and rigidly rotating stars at mass-shed limit ($\Atil=0)$ are also shown. Our results are displayed as solid lines, while the dash-dotted lines correspond to calculations made by \cite{BSS0} for the same equation of state. Following the classification introduced in Paper I, the sequences are of type A for $\Atil=0.5$ or $0.7$, and of type C for $\Atil=1.0$ (see Section \ref{s:mass} for more detail).\label{comparison}}
\end{center}
\end{figure}

\begin{figure*}
\begin{center}
\includegraphics[angle=0,width=.48\textwidth]{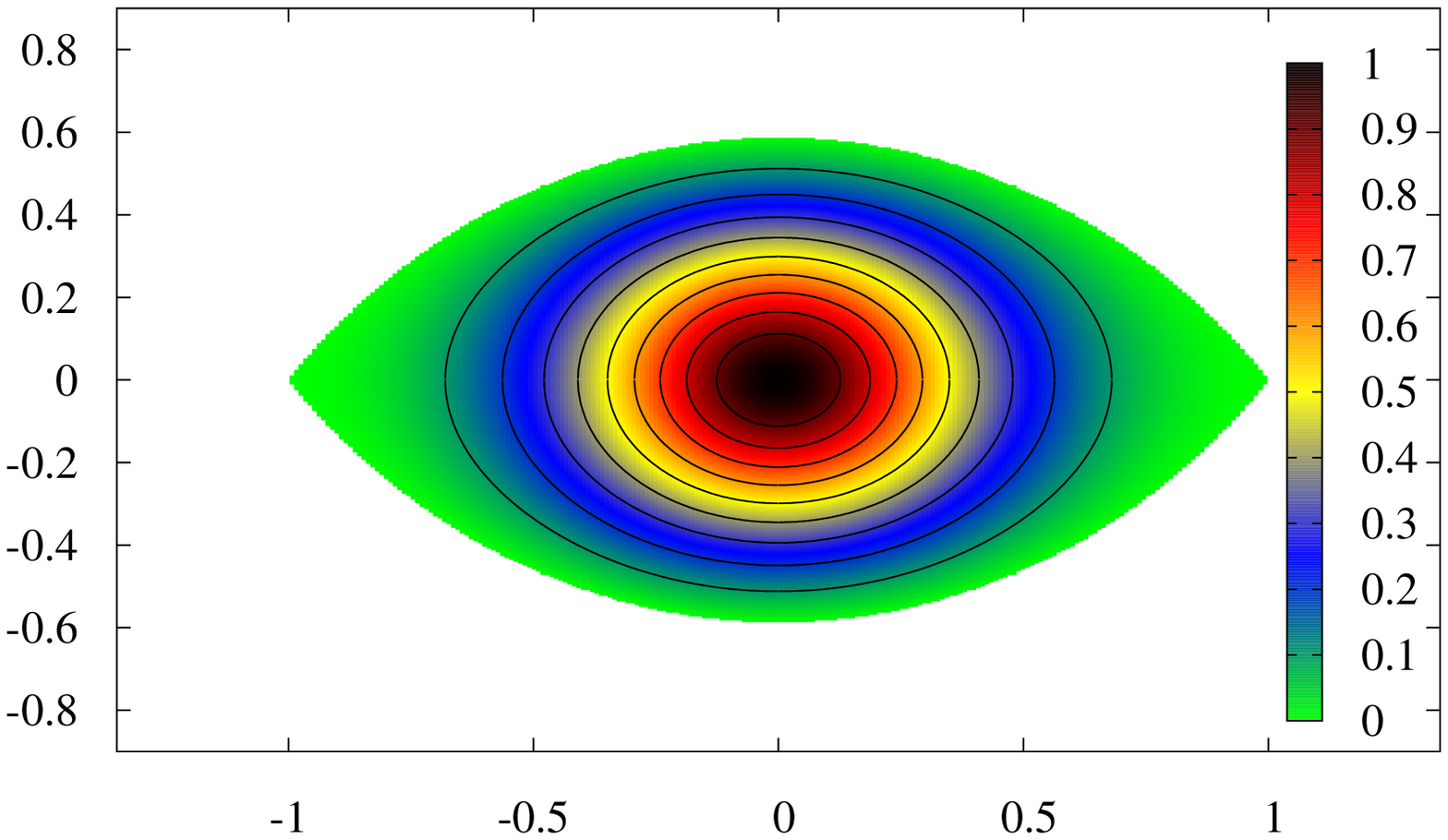}\hspace{.5cm}
\includegraphics[angle=0,width=.48\textwidth]{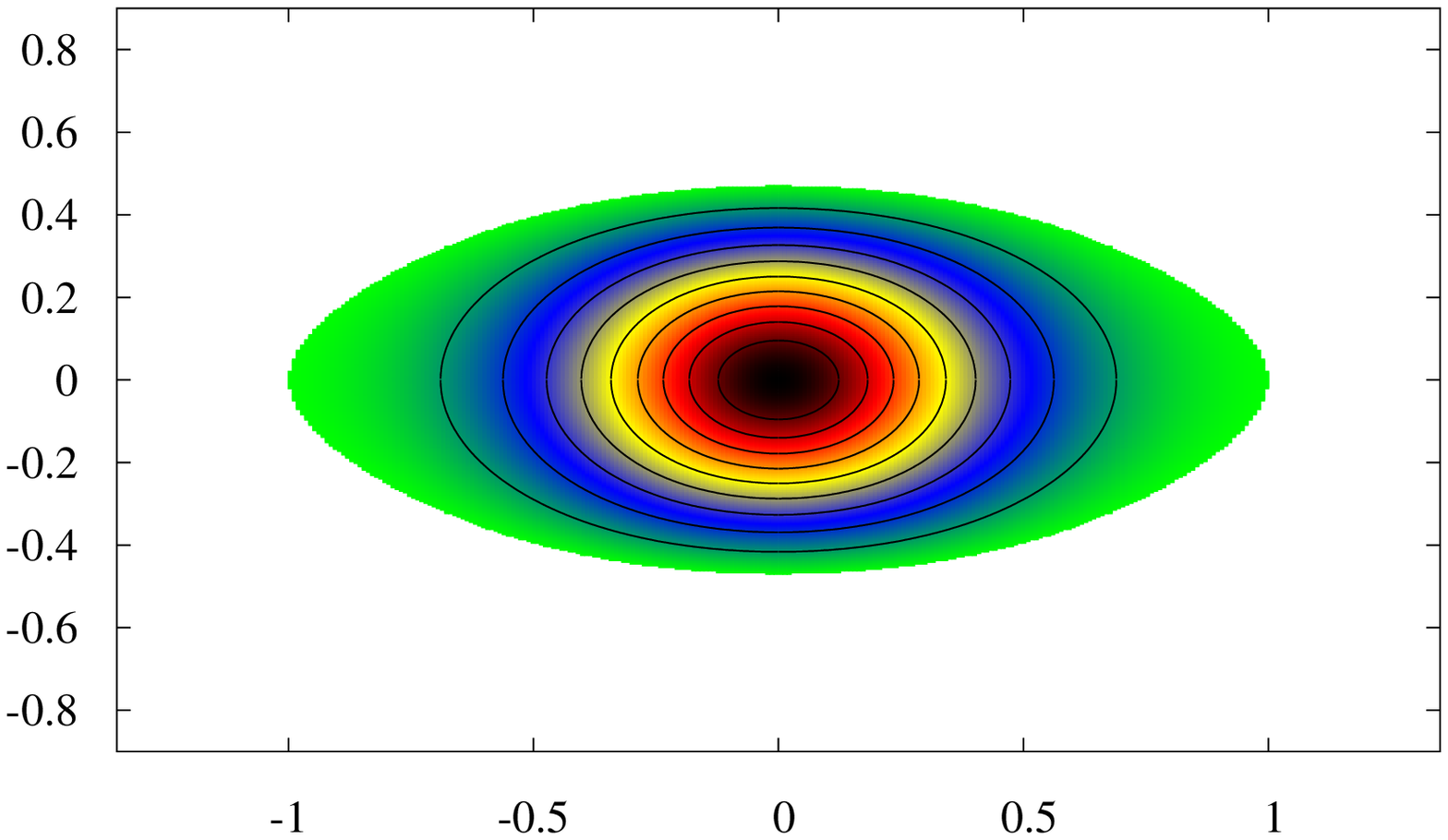}\\
\includegraphics[angle=0,clip,width=0.48\textwidth]{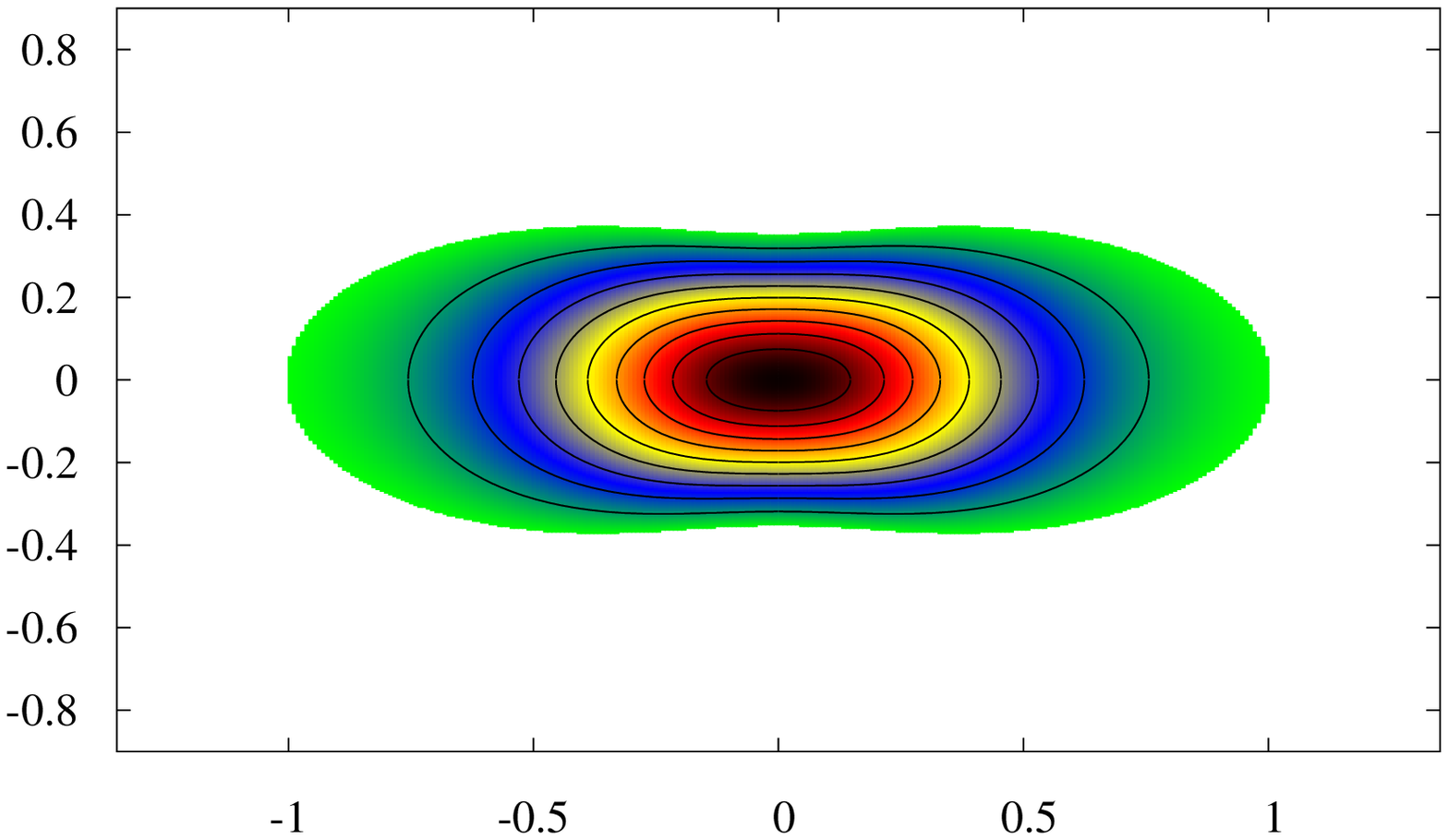}\hspace{.5cm}
\includegraphics[angle=0,clip,width=0.48\textwidth]{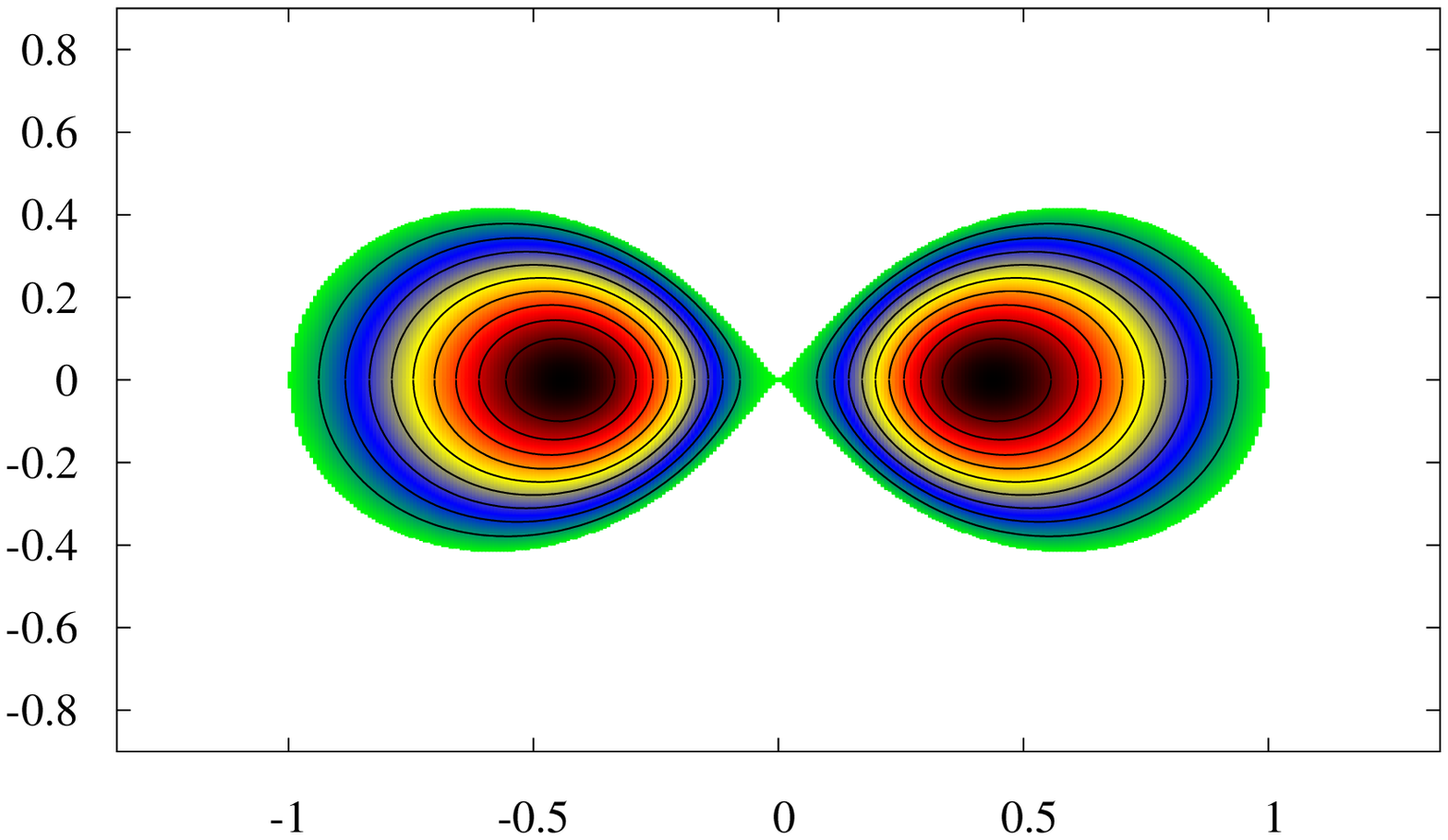}\\
\caption{Isocontours of the relativistic enthalpy $H$ in meridional cross-sections of stars with the maximum allowed mass for rigidly rotating neutron stars at the mass-shedding limit ($\Atil=0$, left upper panel) and three fixed degrees of differential rotation $\Atil=0.5$ (right upper panel, type A), $\Atil=0.7$ (left lower panel, type A) and $\Atil=1.0$ (right lower panel, type C).\label{shapeAC}}
\end{center}
\end{figure*}

 Following \cite{BSS0}, we study the rest-mass~$M_0$ of axisymmetric differentially rotating $\Gamma=2$ polytrops by building sequences of configurations for fixed values of the degree of differential rotation $\tilde A$ and of the maximal energy density $\Em$ in the range $\sim 0$ to $0.6$. Each sequence is parametrized by the ratio of the polar to the equatorial radius $\rrat=r_p/r_e$. We  start with a static, spherically symmetric configuration, $\rrat=1$, and then construct the sequence by decreasing $\rrat$ or, equivalently, by increasing the central angular velocity $\Omega_c$ or the angular momentum $J$, until we reach the mass-shed limit or a star with $\rrat=0$.

As was found in Paper I, for each fixed value of the maximum energy density $\Em$, there is a critical degree of differential rotation $\Atc(\Em)$, below which such a sequence terminates at the mass-shedding limit. Sequences of this kind were assigned the {\bf type A}. On the other hand, when the degree of differential rotation $\tilde A$ is larger than $\Atc(\Em)$, sequences with fixed $\Em$ starting from a static body do not end at the Keplerian limit but enter into the regim of toroidal bodies whose topology is not simply-connected: a hole appears at their center when $\rrat=0$. We call them {\bf type C} sequences. 

For each sequence, be it of type A or C, we looked for the maximum mass. Figure~\ref{typeAC} shows the rest-mass $M_0$ versus $\rrat$ for three examples of sequences with fixed maximal densities $\Em$ and $\Atil$, and for configurations close to the maximum mass. The left panel corresponds to stars with a modest degree of differential rotation $\Atil=0.5$, which end at the mass-shedding limit and are consequently of type A. On the curves, the mass-shedding limit is reached for the smallest non-null value of $\rrat$ that is displayed and that our code always enabled us to calculate. As stated earlier, a configuration at the Keplerian limit is characterized by $\Btil=0$. The vanishing of this rescaled shedding-parameter is evidence of the presence of a cusp at the equator for this kind of configurations. One consequence is that once we have obtained configurations in the neighbourhood of the mass-shedding limit, it is easier, with the Newton-Raphson scheme, to reach the Keplerian limit by decreasing $\Btil$ rather than $\rrat$. As is illustrated by the examples depicted on Fig.~\ref{typeAC}, the maximum mass is found close to but not at the mass-shedding limit (except in the case of uniformly rotating stars). More precisely, we observed that the smaller $\Atil$ is, the closer the configuration with maximum mass is to the mass-shedding limit.

As for the right panel of Fig.~\ref{typeAC}, it shows $M_0$ versus $\rrat$ for type C sequences with $\Atil=1>\Atc$. For such sequences, the maximum mass is always obtained for $\rrat=0$, independly of the value of $\Em$ or of $\Atil$. Notice however that, as in Paper I, we consider in this article only stars without a hole, which implies that we arbitrarily terminate this type of sequences at $\rrat=0$, while the mass can probably become even larger for other non-simply-connected configurations with same values of $\Em$ and $\Atil$.  

In Fig.~\ref{comparison}, we illustrate some of our results by showing the maximum rest mass $M_{0\textrm{max}}$ as a function of the maximum energy density $\Em$ for sequences starting from a non-rotating configuration and with fixed degrees of differential rotation, $\Atil = 0.5$, $0.7$ and $1$. In other words, each curve represents the upper limit on the rest-mass for a given $\Atil$. For those values, lines marked by $\Atil=0.5$ and $0.7$ contain maxima for sequences classified as type A, while for $\Atil=1.0$ they are of type C. We also displayed the results for static stars (lowest line), for rigidly rotating stars at the mass-shedding limit (corresponding to $\Atil=0$) and, for comparison, we included calculations by \cite{BSS0} (dash-dotted lines). It can easily be observed that the agreement with their results is very good for small and moderate degrees of differential rotation $\Atil$ and/or small values of the maximal energy density $\Em$. However, for larger degrees of differential rotation or energy densities, the discrepancy between their results and ours becomes more and more visible. This illustrates that with the high level of accuracy and of stability of our AKM-method based code (see Appendix~\ref{s:NumScheme}), we were able to reach solutions with higher masses than could be considered in previous works based on other algorithms. For each fixed value of the degree of differential rotation $\Atil$, we also indicated on Fig.~\ref{comparison} with a cross the maximum allowed mass (the maximum of maxima).


To illustrate further the differences between the configurations with the maximum allowed mass depending on the degree of differential rotation, we show in Figure~\ref{shapeAC} their shape for $\Atil=0$ (rigid rotation), $0.5$ and $0.7$ (both belonging to type~A sequences), but also $\Atil=1.0$ (a type~C solution). Since for rigid rotation the maximum mass is obtained at the Keplerian limit, the surface of the star is not smooth, but exhibits cusps along the equator. On the contrary, as soon as $\Atil\ne 0$, the maximum mass corresponds to a configuration whose outer surface is regular. As can be seen, the higher $\Atil$, the further from the shape with cusps the maximum mass configuration is. Notably, the configuration with $\Atil=1.0$ has a toroidal shape, but belongs to spheroidal topology (being the last simply-connected object in the type C sequence).


In Table~\ref{tab}, we summarize the properties of differentially rotating stars with maximum allowed mass, for $\Atil$, the degree of differential rotation, ranging from $0$ to $1.5$. For type A solutions, the higher $\Atil$ is, the higher are the allowed mass~$M_{0\textrm{max}}$, the compactness parameter~$M/R_{\rm circ}$ (where $M$ and $R_{\rm circ}$ are the gravitational mass and the circumferential stellar radius, respectively), the angular momentum~$J$, the ratio between the kinetic and the gravitational potential energies~$T/W$, and the Kerr parameter $J/M^2$ (which are indicators of the onset of instabilities for rotating stars, see Section~\ref{s:rotpr}). The higher $\Atil$ is, the lower the maximum energy density is. Note that for configurations with the maximum mass belonging to the type A solution, the maximal density is always located in the stellar center. In contrast, for type C sequences, the maximum allowed mass~$M_{0\textrm{max}}$ is always obtained for $r_{\rm ratio}=0$ and it is a decreasing function of $\Atil$. For such configurations, its highest value is obtained for the smallest possible value of $\Atil$ compatible with type C ($\Atil =\Atc$), which is {$\sim 0.7$} for the $\Gamma=2$ polytropic EOS. However, despite what could have been expected, $M_{0\textrm{max}}$ is not a continuous function of $\Atil$ due to the existence of the different types, and more specifically to the ambiguity of the definition of the types for configurations with $\Atil=\Atc$. This ambiguity, as we shall see in the next subsection, is related to the fact that, as was shown in Paper~I, there are in the solution space other types of sequences than those discussed up to now (types A and C). Furthermore, as can already be seen in Table~\ref{tab}, those types can be associated to even higher masses than the most common type A.

\setcounter{table}{0}
\begin{table*}
\begin{center}
\begin{minipage}{140mm}
\begin{center}
 \caption{Properties of stars with maximal rest-mass~$M_{0\textrm{max}}$ for all types of sequences and for $\Atil$ in $[0; 1.5]$. In addition to the mass, for each configuration are displayed the ratio between the central and equatorial angular velocities ($\Omega_c / \Omega_e$), the ratio between the kinetic and gravitational binding energies ($T/|W|$), the ratio between the polar and equatorial radii ($r_p/r_e$), the maximum enthalpy $\Hm$, the maximum energy density $\Em$ (with the central energy density $\epsilon_c$ in units of $\Em$), the angular momentum $J$, the Kerr parameter $J/M^2$ (with $M$ the gravitational mass) and the compactness parameter $M/R_{\rm circ}$ (with $R_{\rm circ}$ the circumferential radius). For more detail on the accuracy, see Appendix~\ref{s:NumScheme}\label{tab}.}
\begin{tabular}{ccccccccccc}
\hline
\hline\\[-8pt]
TYPE &$\Atil$ & $\Omega_c / \Omega_e$ & $M_{0\textrm{max}}$ & $T/|W|$ & $r_p/r_e$ & $\Hm$ & $\Em\,(\epsilon_c/\Em)$ &$ J $& $J/M^2$ & $ M/R_{\rm circ}$\\
\hline
       &0.0 & 1.000 & 0.206941 & 0.0832 & 0.58479 & 0.43773  & 0.350 (1) & 0.02017 &  0.5690 & 0.17373\\
       &0.1 & 1.027 & 0.207924 & 0.0856 & 0.57966 & 0.43702  & 0.349 (1) & 0.02067 &  0.5773 & 0.17393\\
       &0.2 & 1.108 & 0.210921 & 0.0926 & 0.56477 & 0.43524  & 0.347 (1) & 0.02215 &  0.6014 & 0.17448\\
       &0.3 & 1.240 & 0.216118 & 0.1042 & 0.54132 & 0.43151  & 0.343 (1) & 0.02470 &  0.6389 & 0.17557\\
       &0.4 & 1.422 & 0.223975 & 0.1204 & 0.51059 & 0.42494  & 0.335 (1) & 0.02849 &  0.6871 & 0.17759\\
TYPE A &0.5 & 1.657 & 0.235568 & 0.1419 & 0.47306 & 0.41433  & 0.323 (1) & 0.03406 &  0.7440 & 0.18122 \\
       &0.6 & 1.959 & 0.253800 & 0.1708 & 0.42686 & 0.39772  & 0.304 (1) & 0.04286 &  0.8094 & 0.18834\\
       &0.7 & 2.507 & 0.295169 & 0.2222 & 0.35240 & 0.37329  & 0.306 (1) & 0.06243 &  0.8921 & 0.21434\\
\hline
       &0.8 & 2.999  & 0.46319 & 0.2937 & 0.005 & 0.16416 & 0.097 (2.e-4)& 0.1758 & 1.023 & 0.2504\\
TYPE C &0.9 & 3.382  & 0.43357 & 0.2854 & 0.002 & 0.16750 & 0.100 (2.e-5)& 0.1526 & 1.008 & 0.2450\\
       &1.0 & 3.805  & 0.40851 & 0.2771 & 0.005  & 0.1720 & 0.103 (2.e-4)& 0.1338 & 0.989 & 0.2415\\
       &1.5 & 6.420  & 0.32590 & 0.2379 & 0.01 & 0.1968 & 0.121 (6.e-4) & 0.0783 & 0.897 &  0.2275\\
\hline
\hline
       &0.4 & 1.785 & 0.721 s & 0.336 & 0.035 & 0.152 & 0.089 (0.016) & 0.422 & 1.078 & 0.270\\
TYPE B &0.5 & 2.006 & 0.639 s & 0.335 & 0.114 & 0.145 & 0.084 (0.26) & 0.340 & 1.082 & 0.246  \\
       &0.6 & 2.223 & 0.571 s & 0.331 & 0.144 & 0.140 & 0.081 (0.51) & 0.277 & 1.088 & 0.222\\
       &0.7 & 2.443 & 0.510 s & 0.324 & 0.164 & 0.140 & 0.081 (0.75) & 0.225 & 1.091 & 0.201\\
\hline

TYPE D & 0.8 & 2.6279 & 0.4485 & 0.3116 & 0.1825 & 0.14242 & 0.08239 (0.944) & 0.178 & 1.096 & 0.177 \\

\hline
\end{tabular}
\end{center}
\end{minipage}
\end{center}
\end{table*}

\subsection{Maximum mass for all types of solution}

\begin{figure}
\begin{center}
\vskip 0.5cm
\includegraphics[width=0.4\textwidth,clip,angle=0]{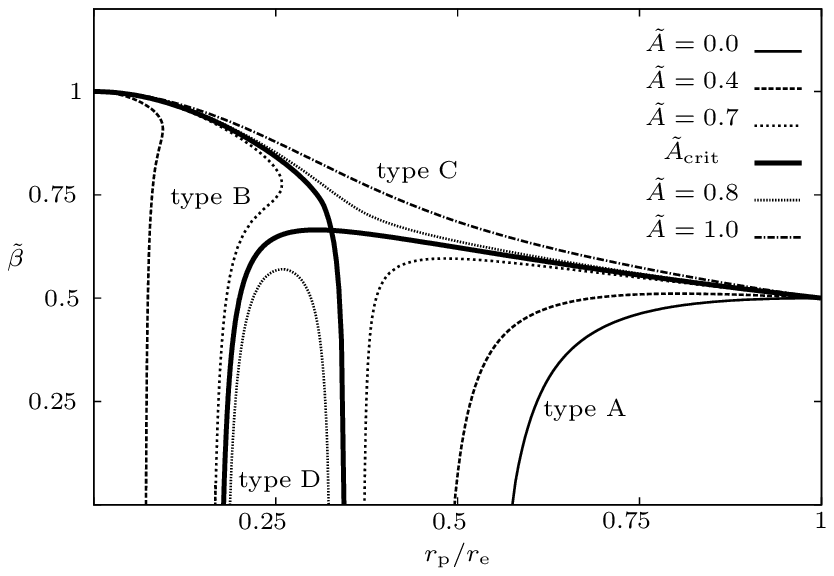}
        \caption{Typical structure of the solution space illustrating, in the $(\rrat,\Btil)$ plane, the various types of sequences for several values of the degree of differential rotation $\Atil$. The curves show the dependency between the shedding parameter $\Btil$ and the ratio between polar and equatorial radii $\rrat$ for $\Gamma=2$ polytropic stars with fixed maximal energy density $\Em=0.12$ ($\Hm=0.2$). The bold curve corresponds to the separatrix sequence with $\Atil=\Atc=0.75904$, which divides the diagram in 4 regions containing sequences of types: A (a lower right corner), B (a lower left corner), C (above separatrix) and D (between types A and B)\label{alltypes}.}
\end{center}
\end{figure}

In Paper~I, it was shown that, for fixed $\Em$ and moderate degree of differential rotation, there are sequences of stars, without a static limit, coexisting either with type A or type C sequences. They belong to new types of sequences, called {\bf type B} and {\bf type D} respectively, and exist only thanks to differential rotation. The four types of one-dimensional parameter sequences, denoted {\bf A, B, C} and {\bf D}, are illustrated in the $(\rrat,\Btil)$ plane for fixed $\Em=0.12$ by Fig.~\ref{alltypes}, on which it can be seen\footnote{We remind the reader that, in such a plane, $(\rrat=1,\Btil=0.5)$ corresponds to a spherical static star, while $(\rrat=0,\Btil=1)$ is the entrance in the toroidal regim and $\Btil=0$ is associated to the mass-shedding limit.} that there are two threshold values of $A$, respectively $\Atil_B$ and $\Atil_D$, such that types A and B coexist for $\Atil_B<\Atil<\Atc$, while types C and D are found when $\Atil_D>\Atil>\Atc$. For a given $\Em$, the minimal value of the degree of differential rotation, $\Atil_B$, for which the type B exist, is the minimum of the $\Atil(\Btil)$ function for fixed $r_{\rm ratio}=0$, and similarly $\Atil_D$ is the maximum of the $\Atil(r_{\rm ratio})$ function for fixed $\Btil=0$. On the other hand, the curve with $\Atil=\Atc$ is a separatrix which divides the plan in four domains. For a given $\Em$, the value of $\Atc$ can be determined thanks to the fact that the $\Atil(\rrat,\Btil)$ function possesses a saddle-point which belongs to the separatrix and at which the four types coexist (see Paper I for details).

Having this in mind the two types of sequences without a static limit, type B and type D, are defined as follows:
\begin{itemize}
\item Type B are one-dimensional sequences that start at the mass-shedding limit ($\Btil=0$) but continously enter into the toroidal regime ($\rrat \to 0$), when fixing $\Em$ and $\Atil$, but varying another suited parameter. As a consequence, they are always characterized by small values of $\rrat$ and, as type C sequences, in our study they arbitrarily end at $(\rrat=0,\Btil=1)$. They are found for $\Atil_B\le\Atil\le\Atc$;
\item Type D also start at the mass-shedding limit ($\Btil=0$) but, unexpectingly, they terminate there as well. As illustrated by Fig.~\ref{alltypes}, configurations of this type fill a smaller part of the solution space and are less easily found than those of all other types. They appear for $\Atil_D\ge\Atil\ge\Atc$.
\end{itemize}

\begin{figure}
\begin{center}
\includegraphics[width=0.48\textwidth,clip,angle=0]{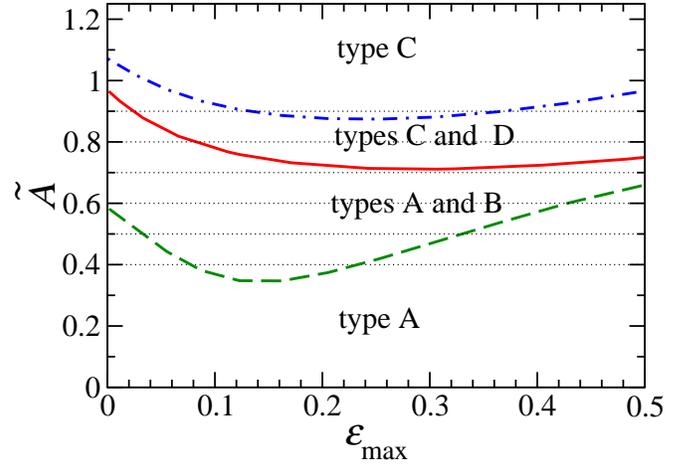}
        \caption{Regions of existence of A, B, C and D types of differentially rotating neutron stars sequences in the plane ($\Atil,\Em$). The central (red) curve indicates the critical value $\Atc(\Em)$ which corresponds to the separatrix, on which all types (A, B, C and D) of solutions coexist. The green dashed line and the blue dash-dotted line correspond to the lower limit of existence of type B, $\Atil_B$, and to the upper limit, $\Atil_D$, of existence of type D, respectively. For $\Atil < \Atil_B(\Em)$ or $\Atil > \Atil_D(\Em)$, only one type remains, respectively A or C\label{regions}.}
\end{center}
\end{figure}

The three threshold values of $\Atil$ are functions of $\Em$: $\Atil_B(\Em)< \Atc(\Em)< \Atil_D(\Em)$. Hence, another useful way to depict the various domains of (co)existence of the types of sequences consist in indicating them in the ($\Em,\Atil$) plane. This is done for the $\Gamma=2$ EOS on Fig.~\ref{regions}, which shows that, for all reasonable values of $\Em$, the solution space has more or less the same structure.

\begin{figure}
\begin{center}
 \includegraphics[width=0.5\textwidth,clip,angle=0]{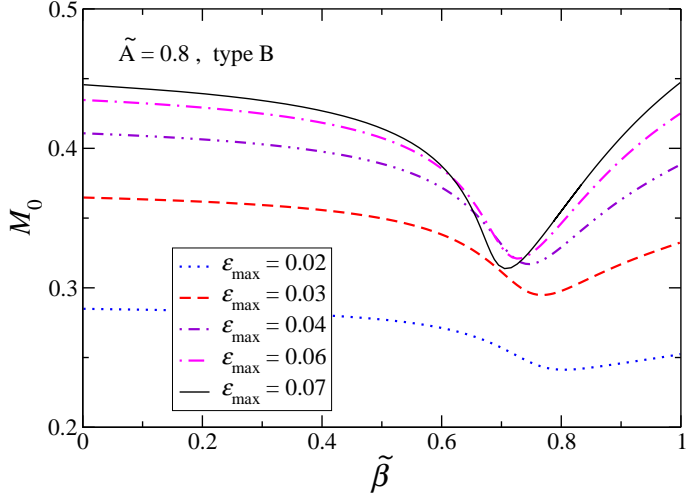}
        \caption{Rest mass $M_0$ as a function of $\Btil$ along sequences of {\bf type B} with fixed $\Em$ (and for all of them $\Atil=0.8$)\label{typeB}.}
\end{center}
\end{figure}

Having understood the global structure of the solution space, we were able to explore it in detail thanks to the high level of flexibility of our Newton-Raphson based code (see Appendix~\ref{s:NumScheme}), which allows to fix or vary any parameter. As can be guessed from Fig.~\ref{alltypes}, the shedding-parameter $\Btil$ is well suited for finding type B sequences with fixed $\Atil$ and $\Em$, at least when we are not too close to the entrance into the toroidal regim $(\rrat=0,\Btil=1)$. Hence, we looked for the maximum mass of this type of sequences by following such curves in the solution space. On Fig.~\ref{typeB}, we show typical results obtained when studying the rest mass $M_0$ as a function of $\Btil$. As explained earlier, all lines start at the mass-shedding limit ($\Btil=0$) and end when entering into the toroidal regim ($\Btil=1$). It was found that for type B, the configuration with maximum mass was always at one or the other of these two positions in the solution space (which can easily be seen for the examples depicted on Fig.~\ref{typeB}). More precisely, when the maximum energy density $\Em$ was sufficiently low, the configuration with maximum mass was at the Keplerian limit, while with increasing $\Em$ it jumped to ($\rrat=0,\Btil=1$). One example of the shape of a star belonging to a type B sequence at the Keplerian limit and with $\rrat \sim 0$ was shown in Fig.~1 of Paper I. It illustrates that the stability of our numerical code allows calculations of extreme configurations simultaneously strongly pinched and oblate.
 
\begin{figure}
\begin{center}
 \includegraphics[width=0.5\textwidth,clip,angle=0]{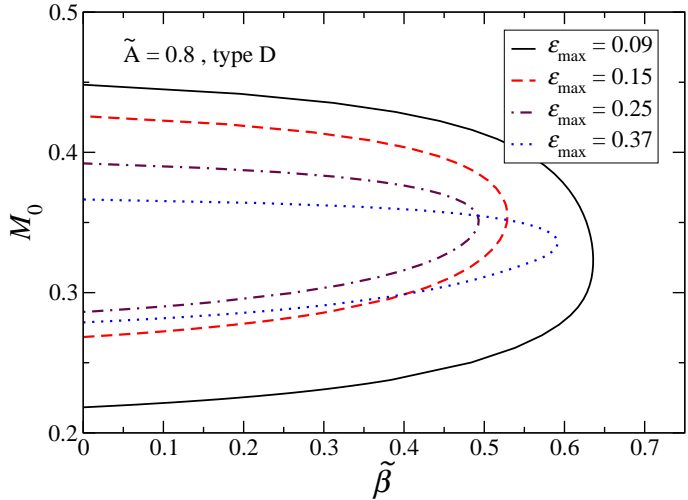}
        \caption{Rest mass $M_0$ as a function of $\Btil$ along sequences of {\bf type D} with fixed $\Em$ (and for all of them $\Atil=0.8$)\label{typeD}.}
\end{center}
\end{figure}

Before commenting further our results, and especially the values of the maximum mass obtained, we shall mention that a quite similar procedure was applied to look for the maximum mass of type D sequences. It led to typical results as those shown by Fig.~\ref{typeD}, which illustrates that the maximum mass was always reached for one of the two mass-shedding configurations belonging to the sequence (more precisely the mass was always for the one with the smallest $\rrat$).

As was already explained, the structure of the solution space is such that, for fixed $\Em$ and $\Atil$, there can be several types of configurations that exist simultaneously. It means that there is not always a unique solution to the system of equations for given values of the three parameters needed to calculate one, depending on what are the chosen parameters. More precisely, we have seen (Fig.~\ref{regions}) that there is some overlap between the domains of existence of types A and B, but also between C and D, if one fixes $\Em$ and varies $\Atil$. Consequently, when looking for the maximum mass for fixed $\Em$ and $\Atil$, as illustrated by curves similar to those on Fig.~\ref{comparison}, all types of sequences have to be taking into account and one should not only consider configurations reached from a static limit. As a matter of fact, when two neutron stars merge, which is one of the situations of interest for studies of maximum masses, a naive expectation could even be that the shape of the material remnant would, at first, be closer to that of a more toroidal type B configuration than to the more spherical shape of a type A star.

\begin{figure}
\begin{center}
 \includegraphics[width=0.5\textwidth,clip,angle=0]{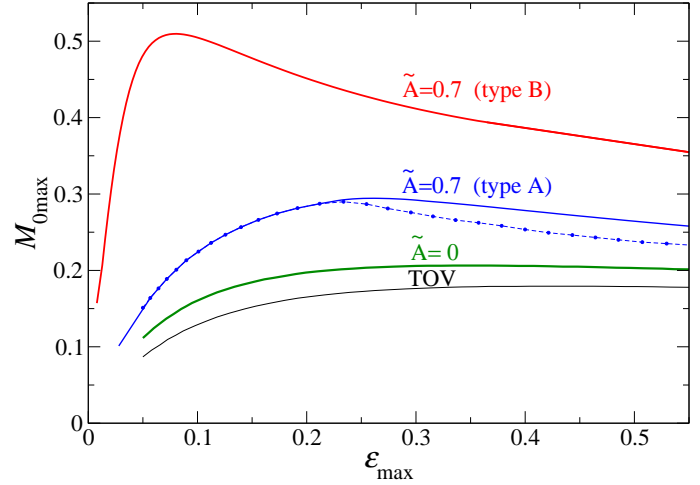}
        \caption{Maximum rest mass $M_{0\textrm{max}}$ as a function of the maximum energy density $\Em$ for $\Atil=0.7$. Curves associated to type A and type B sequences are represented. We also plotted the result for static stars, for rigid rotation ($\Atil=0$), and the data of \cite{BSS0} (dash-dotted line), as in Fig.~\ref{comparison}. One easily observes that stars of type B can be much more massive than those of type A\label{exampleB}.}
\end{center}
\end{figure}

On Fig.~\ref{exampleB}, we come back to some of the results already shown on Fig.~\ref{comparison} and display the maximum rest mass $M_{0\textrm{max}}$ as a function of the maximum energy density $\Em$ for sequences with a fixed degree of differential rotation, $\Atil = 0.7$ [to make easier the comparison we kept the curves for rigid rotation, for static stars as well as the dash-dotted line associated to the calculations of \cite{BSS0}]. However, this time, in addition to the results obtained for type A sequences, we also included those for type B (whose existence for $\Atil=0.7$ is proven by Figs.~\ref{alltypes} and~\ref{regions}).

\begin{figure}
\begin{center}
 \includegraphics[width=0.5\textwidth,clip,angle=0]{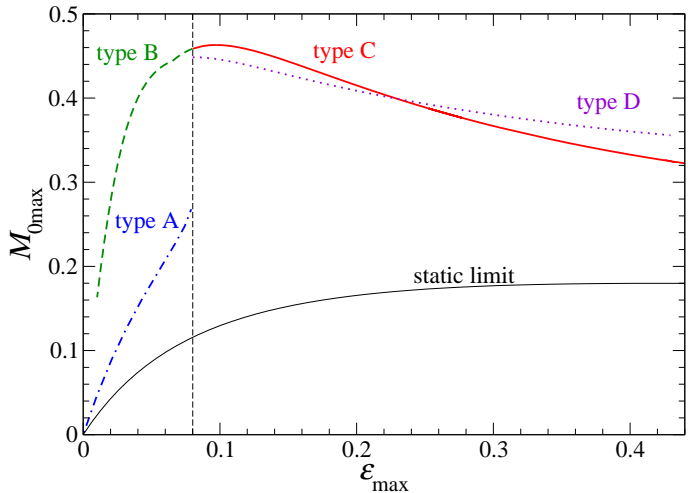}
        \caption{Maximum rest mass $M_{0\textrm{max}}$ as a function of the maximum energy density $\Em$ for $\Atil=0.8$ and all types of solutions\label{A08all}.}
\end{center}
\end{figure}

From this figure, one concludes that type B stars can sustain a much higher mass than the more common type A configurations. If one compares with rigid rotation, the increase of the maximum mass can even be as large as around 150\% (more than $0.5$ compared to $\sim 0.2$). The same kind of conclusions arises from a careful study of the solution space for all types. For instance, Fig.~\ref{A08all} shows the maximum rest mass $M_{0\textrm{max}}$ versus the maximum energy density $\Em$ for $\Atil=0.8$ stars. As can be noticed from Fig.~\ref{regions}, this value of the degree of differential rotation $\Atil$ is one of the few for which all 4 types of solution can exist. More precisely, for low values of $\Em$, one has $\Atc(\Em)>0.8$, so that the configurations are of types A and B, while for large values of $\Em$, $\Atc(\Em)<0.8$, and the configurations are of types C and D, the transition being for $\Em\sim0.08$ such that $\Atc=0.8$ (see Fig.~\ref{regions}).

\begin{figure}
\begin{center}
 \includegraphics[width=0.5\textwidth,clip,angle=0]{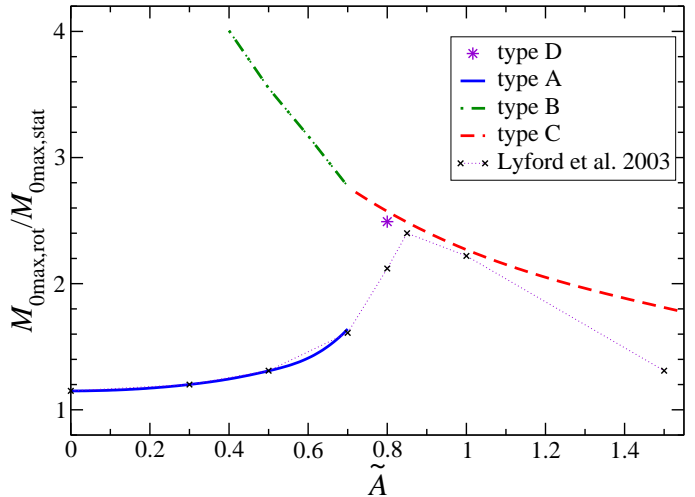}
        \caption{Highest increase of the maximum rest-mass with respect to static configurations as a function of $\Atil$ for all types of sequences. As in Table~\ref{tab}, only one value is given for type D due to the narrowness of its domain of existence (see text for more detail). For comparison, we also display the results of \cite{LBS03}\label{allLBS}.}
\end{center}
\end{figure}

One of the important conclusions that can be drawn from Fig.~\ref{A08all} is that types C and D stars also give rise to masses much larger than those of rigidly rotating stars, even for reasonable degree of differential rotation and maximum energy density. The precise values that we obtained for the highest increase of the maximum rest mass (for given $\Atil$) within sequences of type C and D are presented in Table~\ref{tab}, together with other physical properties of those configurations that we shall describe in Section~\ref{s:rotpr}. Due to the quite small domain of existence of type D, be it in the ($\Em,\Atil$) plane (visible on Fig.~\ref{regions}) or in a ($\rrat,\Btil$) plane with fixed $\Em$ and $\Atil$ (visible on Fig.~\ref{alltypes}), we included only one typical value for this type. Finally, we represented the highest increase of the maximum rest mass with respect to static configurations for all types of sequences on Fig.~\ref{allLBS}, in which we also displayed the results of \cite{LBS03} for comparison. 

To sum up our results, we found that the maximum mass of differentially rotating neutron stars depends on both the degree of differential rotation and the type of solution. From Fig.~\ref{allLBS}, one deduces that the maximum mass is an increasing function of $\Atil$ for type A solution (associated to a low and modest degree of differential rotation), and a decreasing function for types B and C. Furthermore, configurations from the newly discovered types B and C can possess masses much larger than those of the type A. More precisely, the highest increase of the maximal mass, around 4 times the maximal static mass, was obtained for a modest degree of differential rotation and for configurations belonging to type B sequences which were not taken into account in other studies, mainly due to numerical limitations. Stars of type B sequences are indeed very oblate objects, with toroidal shapes such that $\rrat < 0.25$. The agreement of our results with those of \cite{LBS03} is very good for type A solutions, and good for a modest degree of differential rotation, $\Atil=0.8-1.0$, for type C solutions. Our calculations of the maximum allowed mass are consequently the first which take into account all types of solutions. However, the obtained value of the maximum mass is much higher than the mass of the heaviest stars known up to now. It is naturally an open question whether the considered configurations could be stabilized by differential rotation.

\section{Rotational properties}\label{s:rotpr}

For astrophysical purposes, the question of the maximal mass of differentially rotating neutron stars is strongly related to that of their stability. Even without doing dynamical simulations or stability analysis, some basic conclusions can be drawn from the study of rotational properties of the stars, some of which are listed in Table~\ref{tab}. In this Section, we shall focus on some on those properties, restricting the discussion to 
\begin{itemize}
\item{} the comparison of the different types of configurations that all co-exist if $\Atil$ is fixed at $\Atil\,\equiv\,0.8$ (as can be seen on Figure~\ref{regions}), a value of the degree of differential rotation which is an intermediate one;
\item{} their evolution for stars with maximal mass when $\Atil$ changes (see Table~\ref{tab}).
\end{itemize}

\subsection{Angular momentum}

\begin{figure}
\begin{center}
 \includegraphics[width=0.5\textwidth,clip]{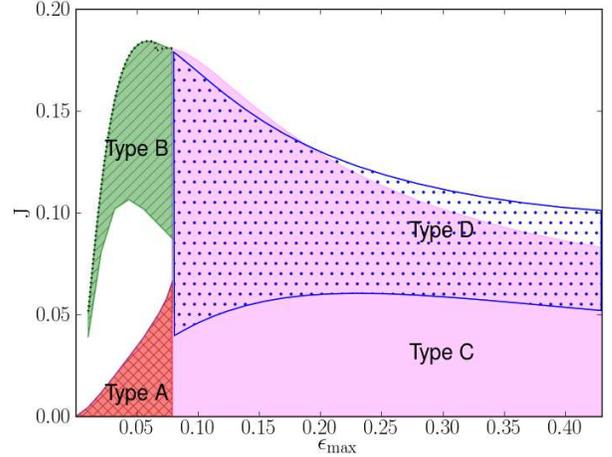}
        \caption{Ranges of angular momentum as a function of the maximal energy density $\Em$ for types A, B, C and D with $\Atil=0.8$. For types A and D, the upper limit always corresponds to the configuration with maximum rest mass. For type C, they nearly coincide, while for type B, it is the case at low $\Em$, but not when one approaches the separatrix. In such conditions, the maximal angular momentum of type B stars is found for the mass-shedding limit whereas the configurations of maximal mass are indicated by a dotted-line on this figure (see text for more detail).}
\label{limitsJ}
\end{center}
\end{figure}

The most obvious quantity to start with is angular momentum $J$, which is depicted on Figure~\ref{limitsJ} as a function of the maximal energy density $\Em$ for stars with $\Atil\,=\,0.8$. A straightforward remark to do is that for types A and C the minimal value of $J$ for fixed $\Em$ is $0$, while it is not for types B and D, which results from the fact that types A and C admit non-rotating limits while B and D do not (see Paper I or Figure~\ref{alltypes}). Then, one can notice that for a given value of $\Em$, the angular momentum stored in a type B star is always larger than in a type A star, as was already the case for the mass. Again, in agreement with what was the situation for the mass, the angular momentum of a type C star of fixed maximal energy density can be both higher or lower than for a type D star.\\

If one no longer fixes the maximal energy density, one notices that the values taken by the angular momentum can be much higher for types B, C or D stars than for type A. Since such high values are reached for very small ratios of the radii and possibly for stars that don't have a non-rotating limit, this implies that calculations made only by accelerating a TOV star shall miss most of those configurations. Also, one can see that when the maximal energy density is close to the transition value which corresponds to the separatrix (see Paper I, Section~\ref{s:mass} and Figure~\ref{regions}), there seems to be a huge gap in the maximal angular momentum for the type A star of highest maximal energy density and for the type C star of smallest maximal energy density, both being equal to $\varepsilon_{\textrm{max,S}\,0.8}$ such that $\Atc(\varepsilon_{\textrm{max,S}\,0.8})=0.8$, see Fig.~\ref{regions}. However, as we briefly discussed in Section~\ref{s:mass}, there is here an ambiguity in the definition of the types linked to the fact that for configurations exactly on the separatrix, type A and C share a branch: on Figure~\ref{alltypes}, it is the right one among the two which are going from the point of intersection of the separatrix lines (in bold) to the mass-shedding limit (the horizontal axis defined by $\tilde{\beta}=0$).

Then, one shall notice that for stars of type A and D (but not B or C), the configurations with the highest value of $J$ are also those with the maximal mass, so that they are (see Section~\ref{s:mass}):
\begin{itemize}
\item{} for type A: close to but not at the mass-shedding limit (exception done of the case of rigid rotation);
\item{} for type D: for the smallest value of $r_\textrm{rat}$ among the two mass-shedding limit configurations (see Figure~\ref{typeD}).
\end{itemize}

As far as types B and C are concerned, the situation is slightly different, maybe due to the fact that those types include stars with a vanishing polar radius (since we decided not to study stars with a hole) which have complicated internal distributions of physical quantities. More specifically, we observed that
\begin{itemize}
\item{} for type B: the maximal $J$ is found at the mass-shedding limit, which is also the maximal mass for maximal energy densities not too close to the transition value (see Section~\ref{s:mass}). On Fig.~\ref{limitsJ}, this explained the continuity between the maximal $J$ of types B and D when $\Em$ increases. Notice that on this figure, the configurations of maximal mass for type B are indicated by a dotted-line;
\item{} for type C: configurations with maximal $J$ are very close to those of maximal mass (at vanishing polar radius, see Section~\ref{s:mass}), but they do not exactly coincide with them.
\end{itemize}

Finally, if one comes back to Table~\ref{tab} to have a glimpse at the influence of $\Atil$ on $J$, one can see that increasing the degree of differential rotation ({\it i.e.} increasing $\Atil$) allows higher maximal values of $J$ only for stars of type A. For types B and C, it is the opposite, as was already the case with the maximal mass (see Section~\ref{s:mass}).

\subsection{Kinetic energy to gravitational energy ratio~($T/|W|$) and instabilities}

\begin{figure}
\begin{center}
 \includegraphics[width=0.5\textwidth,clip]{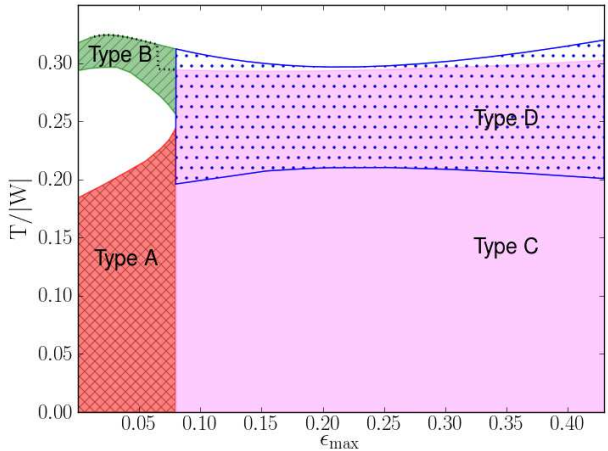}
        \caption{Ranges of the ratio between the kinetic and the potential energies as a function of the maximal energy density $\Em$ for types A, B, C and D with $\Atil=0.8$. Notice that for type B and the highest $\Em$, configurations of maximal $T/|W|$ and maximal mass do not coincide. The latter appear in this figure as a dotted-line. Again, there is no gap between the lines of maxima for types B and D since on the separatrix they coincide at the mass-shedding limit of lowest $r_\textrm{rat}$ (see text for more detail).}
\label{TW}
\end{center}
\end{figure}

For observational purposes, a quantity which is more directly interesting than the angular momentum is $T/|W|$, the ratio between the kinetic (rotational) energy $T$ and the gravitational binding energy $W$. This ratio indeed plays the role of an order parameter \citep{BR76} and is a good indicator of the possible onset of instabilities that can be source of gravitational waves \citep{And2003}. It is displayed on Fig.~\ref{TW} for stars with $\Atil\,\equiv\,0.8$.\\

This figure shows many similarities with the figure for angular momentum $J$ (Fig.~\ref{limitsJ}), for instance in the facts that 
\begin{itemize}
\item{} type B allows for higher values than type A;
\item{} for types A and C the minimal value if 0, while it is not for types B and D;
\item{} there seems to be a discontinuity of the maximal value when one goes from type A to type C, which is again due to the ambiguity in the definition of the types on the separatrix;
\item{} for types A and D, the maximal value of $T/|W|$ is obtained for the configuration with maximal mass, whereas for type C they almost coincide and for type B it happens only for the lowest $\Em$.
\end{itemize}

However, one shall also notice some changes with respect to the situation for $J$:
\begin{itemize}
\item{} the relative difference between the possible values for type A and for other types is not as large for $T/|W|$ as it was for $J$;
\item{} type C does not allow larger values than type D;
\item{} for all types, the maximal value depends much less on $\Em$ than it did for $J$.
\end{itemize}

As far as the influence of the degree of differential rotation is concerned, Table~\ref{tab} tells us that in a way similar to what happens for $J$ and $M_0$, the maximal value of $T/|W|$ is an increasing function of this degree ({\it i.e.} of $\Atil$) only for stars of type A, and it is a decreasing one for types B and C. Of course, this table also confirms that much higher values of $T/|W|$ can be reached for types other than A.\\

\begin{figure}
\begin{center}
 \includegraphics[width=0.5\textwidth,clip]{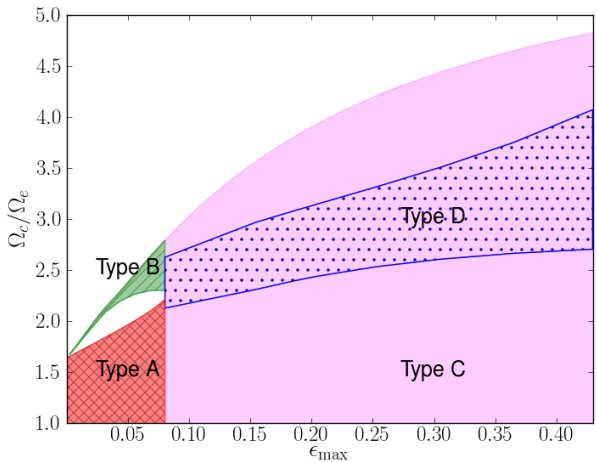}
        \caption{Ranges of the ratio between the central and equatorial angular velocities $\Omega_c/\Omega_e$ as a function of the maximal energy density $\Em$ for types A, B, C and D with $\Atil=0.8$.}
\label{Omega}
\end{center}
\end{figure}

Indeed, another result illustrated by Fig.~\ref{TW} is that for all configurations of types B and D, $T/|W|$ is at least equal to $0.2$, which leads to the legitimate question of the (dynamical) stability of such stars [see \citet{SBS0,And2003}]. The study of this stability is beyond the scope of this work since it requires making dynamical simulations \citep{BSS0,SBS0} or analysis of perturbations, but refering to previous studies, one can expect such dynamical instabilities as the so-called bar-mode instability. Furthermore, as we are dealing here with differentially rotating stars, another kind of instability could be triggered, the so-called low $T/|W|$ instability [see for instance \citet{KGK10} and references therein]. As a relation between the later and the appearance of a co-rotation point has been suggested [see \citet{PA15} and references therein], an easy way to get some more information on the possible stability of the configurations under study here is by depicting the ratio between central and equatorial angular velocities, $\Omega_c/\Omega_e$, as is done on Fig.~\ref{Omega} for stars with $\Atil=0.8$.\\

Again, the picture is quite similar to the previous ones, with 
\begin{itemize}
\item{} type B which corresponds to higher values than type A;
\item{} types A and C which have 0 as a minimal value whereas types B and D do not;
\item{} an apparent discontinuity of the maximal value when one goes from type A to type C, which is still due to the ambiguity in the definition of the types on the separatrix.
\end{itemize}

However, the ratio $\Omega_c/\Omega_e$ differs from the other quantities by the fact that, for types B and D, its maximal value is obtained for the configuration with maximal mass, while for types A and C they only almost coincide. Notice that the value of this ratio for the configuration of maximal mass is so close to its maximal value that we do not indicate it on Fig.~\ref{Omega}.

As far as the quantity $\Omega_c/\Omega_e$ itself is concerned, this figure shows that types B, C and D (and especially type C) allow large values (up to 5 for $\Atil=0.8$, a moderate degree of differential rotation), which implies a large window of possible corotation for instabilities such as those studied in \citet{KGK10} to be triggered. As we stated earlier, we shall not enter more into the detail of this topic, and we just notice to conclude that Table~\ref{tab} displays, as one can expect, a strong and positive correlation between the value of $\Omega_c/\Omega_e$ for the star with maximal mass and the value of $\Atil$.

\subsection{Kerr parameter $J/M^2$}

\begin{figure}
\begin{center}
 \includegraphics[width=0.5\textwidth,clip]{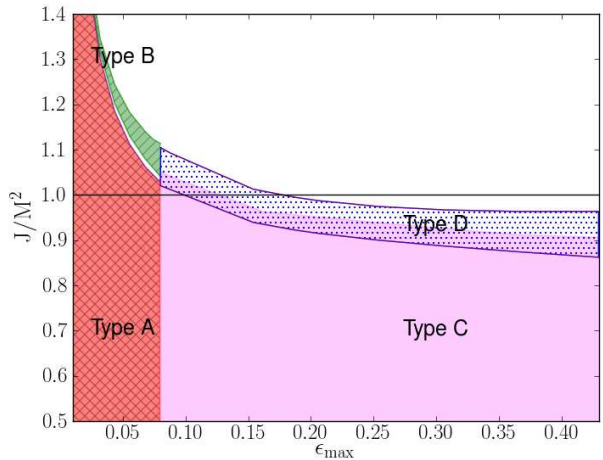}
        \caption{Ranges of the ratio between the angular momentum and the square of the gravitational mass, $J/M^2$, the so-called Kerr parameter which cannot be larger than $1$ (limit indicated by the black horizontal line) for a rotating black hole in general relativity. It is displayed here as a function of the maximal energy density $\Em$ for types A, B, C and D with $\Atil=0.8$.}
\label{Kerr}
\end{center}
\end{figure}

To conclude our brief study of rotational quantities for stars with maximal mass or with a rotation profile characterized by $\Atil=0.8$, we shall look at the so-called Kerr parameter $J/M^2$ whose value cannot be larger than~$1$ for a rotating black hole in general relativity. Indeed, it has been suggested [see for instance \citet{GRS11}] that supra-Kerrian stars whose collapse would lead to a naked singularity are somehow stabilized so that the cosmic censorship conjecture is respected. As previously done for other rotational quantities, we picture on Fig.~\ref{Kerr} the Kerr parameter $J/M^2$ as a function of $\Em$ for stars from all types and with $\Atil=0.8$, and we show this parameter for all stars with maximal mass but various values of $\Atil$ and types in Table~\ref{tab}.

From the table or the figure, it can easily be checked that as soon as the density is large enough, no supra-Kerrian configuration exist, a conclusion quite similar to what was found in \cite{GRS11}. More precisely, for types A, B and C, the Kerr parameter is a decreasing function of $\Em$. Additionally, we observe from Table~\ref{tab} that for stars with maximal mass, $J/M^2$ is an increasing function of the degree of differential rotation ($\Atil$) for types A and B, whereas it is a decreasing one for type C. Furthermore, the Kerr parameter is always larger than~$1$ for type B stars and it can be so for type D stars whose maximal density is not too large (such as the star with maximal mass displayed in Table~\ref{tab}). Although we previously saw that quantities such as $T/|W|$ or $\Omega_c/\Omega_e$ strongly suggest that they are not stable, the fact that the Kerr parameter is larger than~$1$ for such configurations could be a possible indication of their quasi-stability. Indeed, the dynamical simulations of \cite{GRS11} showed that supra-Kerrian stars seem to be stabilized by differential rotation. If other conclusions from \cite{GRS11} are correct, the final collapse of such stars would be associated with the excitation of various modes, making them very interesting sources of gravitational waves. Nevertheless, as was already stated, to be properly dealt with, this issue would require some dynamical study or some perturbative analysis which are far beyond the scope of the current work.

\section{Discussion of the results and conclusion}\label{s:conc}

Using a highly accurate spectral code based on the Newton-Raphson scheme, we calculated configurations of relativistic differentially rotating neutron stars modeled as $\Gamma=2$ polytrops for broad ranges of maximal densities and of the degree of differential rotation. We were able to fully explore the solution space for stars with a rotation profile described by the law proposed in \cite{KEH}, although we considered only models with spheroidal topology (without a hole).

For the first time, the maximum mass and various other astrophysical quantities were calculated for all types of differentially rotating neutron stars, as defined in Paper~I. The maximum mass of differentially rotating neutron stars was shown to depend not only on the degree of differential rotation but also on the type of the solution. Its value is an increasing function of the degree of differential rotation for type A solutions (associated to a low or to a modest degree of differential rotation) and a decreasing function for types B (with a modest degree of differential rotation) and C (with a modest or a high degree of differential rotation). The highest increase of the maximal mass, 3-4 times the maximal non-rotating mass, is obtained for intermediate degrees of differential rotation, indicating that the corresponding configurations could be relevant in some astrophysical scenarios. Those configurations belong to sequences of type B which were not taken into account in previous studies mainly due to numerical difficulties. In addition, the thorough investigation performed in the present article allowed to understand the partial results obtained with other codes \citep{BSS0,LBS03} and to show that the maximum possible mass of a differentially rotating neutron star could be much higher than previously thought, even for astrophysically reasonable configurations.

In order to try to go a step further in deciding whether configurations of the new types have pertinence in actual situations, we performed a rough analysis of their rotational parameters, such as their angular momentum and other quantities linked to the possible appearance of instabilities. We observed that the ratio between the kinetic and potential energies was indeed quite large for many newly discovered configurations, but we also noticed that so is their Kerr-parameter (always higher than 1 for stars with the maximum mass belonging to types B and D and for some of type C), which could imply that they are somehow stabilized. However, the definitive answer to that question has to come from other analysis, be they perturbative or fully dynamical. A few hydrodynamical studies \citep{BSS0,SBS0,GRS11} have already shown that supra-Kerr stellar models seem dynamically stable but are subject to various secular instabilities leading to the emission of gravitational waves. Another complementary approach would naturally be to use the configurations we have calculated as initial data to perform dynamical evolutions of differentially rotating neutron stars and to study the stability criteria for such objects.

To be of physical interest, the conclusions drawn in our study should as well be supported by further investigations with more relatistic descriptions of the microphysics, such as the equation of state. In \cite{SKGVA}, we present results concerning the influence of the stiffness of a polytropic equation of state on the various types of configurations and on their properties to examine how robust our results are. In other articles, we shall study the maximum mass of strange stars \citep{SGVA} and analyse in detail the rotational properties of neutron stars described by realistic EOSs.


\section*{Acknowledgments}
This work was partially supported by the Polish Grant N N203 511238; the FOCUS Programme of Foundation for Polish Science; by POMOST/2012-6/11 Program of Foundation for Polish Science co-financed by the European Union within the European Regional Development Fund, by the grant of the National Science Centre
UMO-2014/14/M/ST9/00707, UMO-2013/01/ASPERA/ST/0001 and
DEC-2013/08/M/ST9/00664, by the NewCompStar COST Action MP1304 and by
the HECOLS International Associated Laboratory programme. Calculations
were performed on the PIRXGW computer cluster funded by the Foundation
for Polish Science within the FOCUS program.

\appendix

\section{The numerical scheme} \label{s:NumScheme}

As in Paper I \citep{AGV}, the numerical calculations are done using a pseudospectral collocation point method that utilizes two domains: (i) a domain which covers the fluid's interior, and (ii) a spatially compactified domain describing the fluid's exterior. In order to avoid Gibbs phenomena, we choose the common boundary between the two domains to coincide with the surface shape of the fluid configuration. This shape is not known {\em a priori} but forms part of the elliptic `free' boundary value problem to be solved. Each of the two subdomains is characterized by a mapping

\beq\label{GenMap}
        \varrho^2 = \varrho^2_k(s,t)\,, \qquad z^2=z^2_k(s,t)\,,\quad (s,t) \in [0,1]^2
\eeq
where $\varrho$ and $z$ are the coordinates used in the metric (see Paper~I) and where $k$ labels the subdomains ($k\in\{0;1\}$). Here we have used the fact that the solutions are axially and equatorially symmetric from which it follows that the metric coefficients are functions of the coordinate squares, $\varrho^2$ and $z^2$. 

\begin{figure}
\begin{center}
        \includegraphics[scale=0.3,clip]{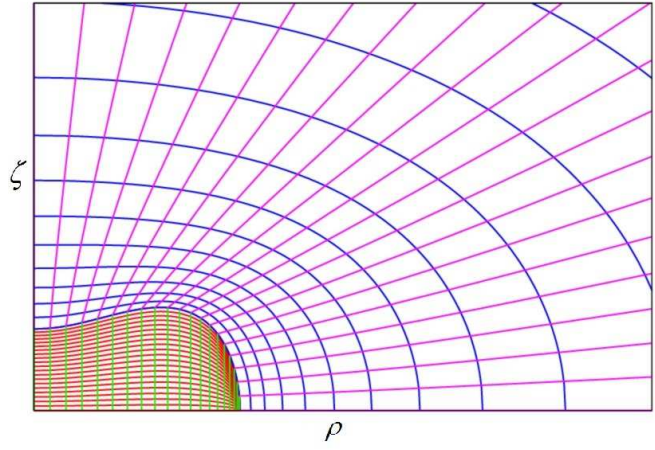}
        \caption{Example for mappings of interior and exterior domains, see (\ref{GenMap}). The coordinate transformations being chosen are specifically suited to extremely flattened configurations.}
\label{Coordinates}
\end{center}
\end{figure}

For the coordinate transformation
\[
        (s,t) \mapsto (\varrho^2, z^2)
\]
we take care of the fluid's unknown surface shape by means of a one-dimensional function $G$,
\[
        G:[0,1]\to \mathbb{R}\,.
\]
In particular we write:
\begin{itemize}
        \item Exterior subdomain, $k=0$:
                \bea
                        \varrho_0^2(s,t) &=& t\left[r^2_\mathrm{e}-r^2_\mathrm{p}+\xi^2(s)\right] \\
                        z_0^2(s,t)   &=& (1-t)\left[\xi^2(s)-r^2_\mathrm{p}\right] + G(t)-r^2_\mathrm{e}t
                \eea
        with
                \bea
                        \xi(s)    &=& r_\mathrm{p}+r_\mathrm{e}\frac{1-\sigma(s)}{\sigma(s)}\,, \\
                        \label{rescale_s}\sigma(s) &=& 1-\frac{\sinh[(1-s)\log\varepsilon_s]}{\sinh(\log\varepsilon_s)}\,,
                \eea
                \bea
                        G(t=0)    &=& r^2_\mathrm{p}\,, \\
                        G(t=1)    &=& r^2_\mathrm{e}\,.
                \eea
        Here, $r_\mathrm{p}$ and $r_\mathrm{e}$ describe the polar and equatorial radii respectively.

        The boundaries of the exterior domain are described by:
                \bea
                        s&=&0:\,\mbox{Spatial infinity, $\sqrt{\varrho^2+z^2}\to\infty$} \\
                        s&=&1:\,\mbox{Surface of the fluid, given by}:\label{SurfaceDescription} \\ 
                           && \qquad \left\{(\varrho^2,z^2)=
                                \left(r_\mathrm{e}^2t,[G(t)- r^2_\mathrm{e}t]\,\right), t\in[0,1]\right\}\\
                        t&=&0:\,\mbox{Rotation axis, $\varrho=0$}\\
                        t&=&1:\,\mbox{Equatorial plane, $z=0$}
                \eea
        \item Interior subdomain, $k=1$:
                \bea
                        \varrho_1^2(s,t) &= & r^2_\mathrm{e}t \\
                        z_1^2(s,t)       &= & s\left[G(t)- r^2_\mathrm{e}t\right]\,.
                \eea
        The boundaries of the interior domain are described by:
                \bea
                        s&= &0:\,\mbox{Equatorial plane, $z=0$}\\
                        s&= &1:\,\mbox{Surface of the fluid, as in (\ref{SurfaceDescription})} \\ 
                        t&= &0:\,\mbox{Rotation axis, $\varrho=0$}\\
                        t&= &1:\,\mbox{Equator, $\varrho=r_\mathrm{e}, z=0$}
                \eea
\end{itemize}
The mapping of the exterior subdomain is chosen to resemble oblate spheroidal coordinates in which the entire class of Maclaurin spheroids exhibits a rapid spectral convergence rate. For general highly flattened relativistic stars we find, however, that the spectral convergence rate can be improved considerably by refining the exterior spectral mesh in the vicinity of the fluid's surface. We achieve this by rescaling the coordinate $s$ and adjusting the free parameter $\varepsilon_s$ introduced in Eq~(\ref{rescale_s}). For an illustrative example see Fig.~\ref{Coordinates}.

In our pseudospectral collocation point scheme, all functions $U^{\kappa}$
($\kappa=0\ldots n_{\mathrm{eq}}-1$) to be determined by the free boundary value problem are considered at specific gridpoints $(s_{k,i};t_{k,j})$ in the subdomains $k\in\{0;1\}$. These gridpoints are given through:
\bea
        s_{k,i} &=& \sin^2\left(\frac{\pi i}{2(n_k^{(s)}-1)}\right)\;;\quad i=0\ldots n_k^{(s)}-1\\
        t_{k,j} &=& \sin^2\left(\frac{\pi j}{2(n_k^{(t)}-1)}\right)\;;\quad j=0\ldots n_k^{(t)}-1
\eea
The integers $n_k^{(s)}$ and $n_k^{(t)}$ describe the number of gridpoints in the domain $k$ with respect to the $s$- and $t$-directions (\emph{i.e.} the spectral expansion orders). While $n_0^{(s)}$ may be different from $n_1^{(s)}$, we assume the same numbers $n_0^{(t)} = n_1^{(t)}$ of gridpoints at the common domain boundary.

We collect all function values
\beq
        \label{U_values}
        U_{k,ij}^{\kappa} = U^{\kappa}(s_{k,i}, t_{k,j}) \eeq as well as the
        values of the unknown surface function~$G, G_j=G(t_{1,j})$, in order
        to build up a vector ${\bf f}$. In addition, this vector is filled
        with two physical parameters that characterize, for a given equation
        of state, the configuration.  In particular, we choose them to be
        $V_c$ [value of $V$ at the center, see definition (\ref{e:defV})] and
        $\Omega_c$ [see definitions (\ref{e:lawdif1}) or (\ref{e:lawdif2})].
        Note that it is sometimes possible to find more than one solution to a
        given pair of parameters.

The collection of elliptic equations valid in the subdomains, transition conditions at the common domain boundary, the vanishing pressure boundary condition at the fluid's surface and certain parameter relations that one wishes to fulfill, yield a discrete non-linear system of the form 
\beq\label{F_of_f}
        {\bf F^{(n)}(f^{(n)})} = 0
\eeq
where $n$ stands for the collection of all $n_k^{(s)}$ and $n_k^{(t)}$,
\[
        n=\{(n_k^{(s)},n_k^{(t)}); k=0;1\} \;.
\]
 The dimension of this system is given by
\beq
        \label{total_dim}
        n_{\mathrm{total}} = n_{\mathrm{eq}}\sum_{k=0}^{1} n_k^{(s)} n_k^{(t)} + n_\mathrm{G} + n_\mathrm{par},
\eeq 
with $n_\mathrm{G}=n_0^{(t)}$ and $n_\mathrm{par}=2$. In particular, the transition conditions require the $U^\kappa$ to be continuous and to possess continuous normal derivatives. At domain boundaries which correspond to portions of the rotation axis or the equatorial plane, we require regularity conditions, which follow from the elliptic equations when specialized to this boundary. Via the integrated Euler equation (\ref{e:euler}), the vanishing pressure boundary condition restricts the potentials at the fluid's surface. It adds $n_\mathrm{G}$ equations to the system. Finally, we may include specific parameter relations that we wish to be satisfied. For example, we could just prescribe certain values for the physical parameters contained in ${\bf f}$. However, we also might wish to prescribe other parameters instead, say rest mass $M_0$ and angular momentum $J$ of the objects. For this reason we include the $n_\mathrm{par}$ physical parameters into the vector ${\bf f}$ and add the specific parameter relations to the system.

The solution ${\bf f}^{(n)}$ of the discrete algebraic system (\ref{F_of_f}) describes the spectral approximation of the solution to the free boundary value problem. We find the vector ${\bf f}^{(n)}$ using a Newton-Raphson scheme,
\bea\label{NewtRaph}
        {\bf f}^{(n)} &=& \lim_{m\to\infty}{\bf f}^{(n)}_m\,,\\
        {\bf f}^{(n)}_{m+1} &=& {\bf f}^{(n)}_m - \left[{\bf J}^{(n)}({\bf f}^{(n)}_m)\right]^{-1} 
                {\bf F}^{(n)}\left({\bf f}^{(n)}_m\right)\,,
\eea
where the Jacobian matrix is given by
\beq\label{Jacobian}
        {\bf J}^{(n)} = \frac{\partial{\bf F}^{(n)}}{\partial{\bf f}^{(n)}} \,.
\eeq
Note that for the convergence of the scheme a `good' initial guess ${\bf f}^{(n)}_0$ is necessary which we provide through a known nearby function. 

The linear step inside the Newton-Raphson solver, {\it i.e.} the solution of 
\[
        {\bf J}^{(n)} \cdot \delta f = -{\bf F}^{(n)}\;,        
\]
is performed with the preconditioned `Biconjugate Gradient Stabilized (Bi-CGSTAB)' method \citep{Barrett93}. A good convergence of this method requires a so-called preconditioning, which we construct in complete analogy to \cite{Ansorg2004} and \cite{Ansorg2005} through a second or fourth order finite difference representation of the Jacobian matrix of the non-linear system.

\begin{figure}
\begin{center}
 \includegraphics[width=0.6\textwidth,clip,angle=0]{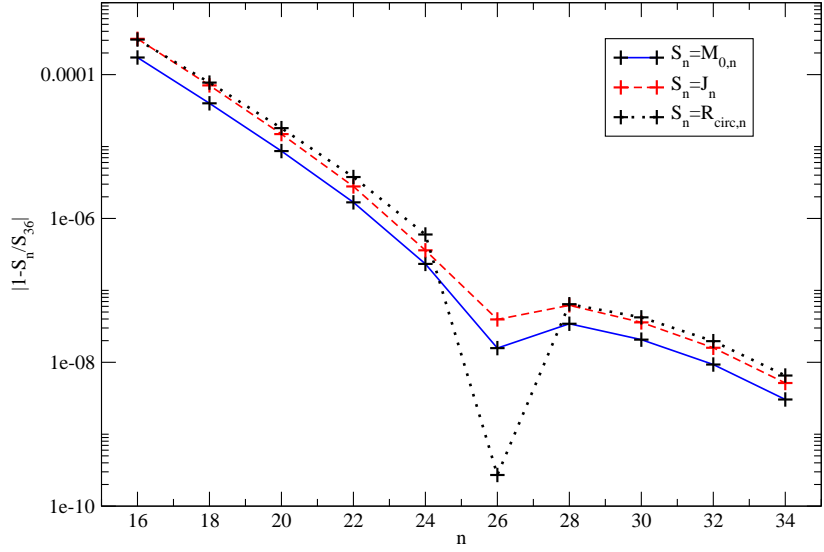}
        \caption{Illustration of the geometrical convergence rate for the rest mass $M_0$, the angular momentum $J$ and the circumferential radius $R_{\rm circ}$ of a differentially rotating neutron star described by a polytropic EOS with $\Gamma=2$. This configuration is of type A and characterized by a degree of differential rotation $\Atil=0.7$, a ratio of the coordinate radii $\rrat= 0.36$ and a maximal enthaly $\Hm=0.38$. For all three quantities, the plot displays the accuracy of the $n$th spectral approximation~$S_n$, with here $n=n_0^{(s)}=n_1^{(s)}=n_0^{(t)}=n_1^{(t)}$\label{convD1}.}
\end{center}
\end{figure}

For most configurations considered in this article, numerical solutions with extremely high accuracy were obtained with a moderate computational effort. This is illustrated by Fig.~\ref{convD1} which displays the accuracy reached for some astrophysical parameters, e.g. the baryon mass, the circumferencial radius and the angular momentum, for a typical example of a not too oblate ($\rrat=0.36$) type A star with a moderate degree of differential rotation ($\Atil=0.7$) and a modest maximum enthalpy ($\Hm=H_{\rm c}=0.38$). More precisely, on this figure, are represented the relative differences between the $n$th spectral approximant $S_n$ of all these quantities (denoted $S$) and their approximant of order $36$, as functions of the number $n$ of spectral points. In the case of more extreme configurations (e.g. close to the Keplerian limit, $\Btil=0$, or to the entrance in the toroidal regime, $\rrat=0$), the number of points needed to get a similar accuracy was larger, but the code also appeared able to perform the calculations. For subcritical configurations, $n=24$ was sufficient, while for extreme ones, up to $n=34$ was sometime necessary. On Fig.~\ref{convD1} can be observed that the accuracy reached for such resolutions is of the order of $10^{-10}-10^{-7}$.

\begin{figure}
\includegraphics[width=0.44\textwidth]{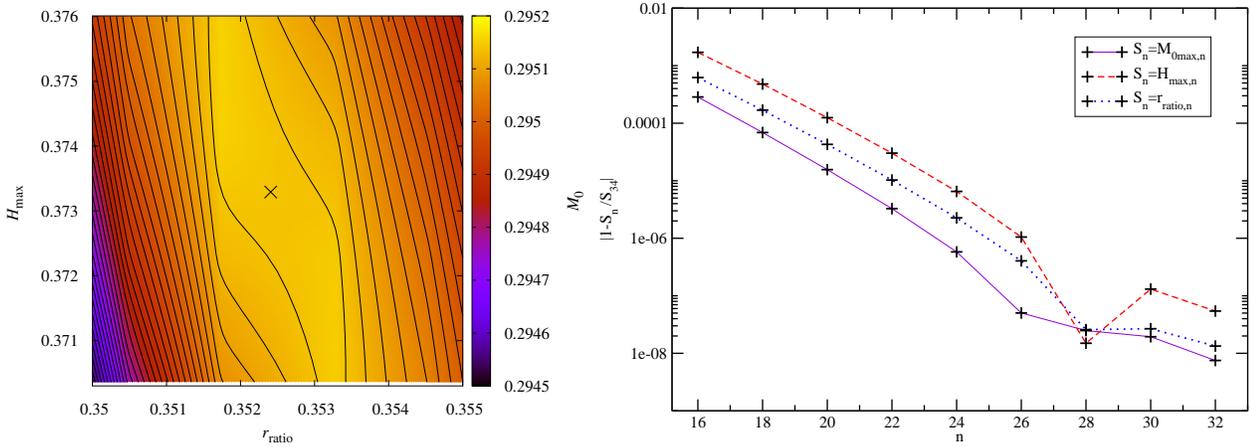}
\includegraphics[width=0.48\textwidth]{Mmax_A07acc_7x7.eps}
        \caption{Left panel: Illustration of the method used and of the precision reached when localizing the configuration with maximum mass for type A sequences with $\Atil=0.7$ (as shown in Table~\ref{tab}). Right panel: geometrical convergence rate for the maximum rest mass $M_0$ and for the corresponding maximum enthalpy $\Hm$ and ratio between the polar and equatorial radii $\rrat$. More specifically, the plot displays the relative accuracy of the $n$th spectral approximations with respect to the $34$th\label{convD2}.}
\end{figure}

To conclude this Appendix on the numerical scheme, we shall briefly describe the method used and the precision reached to identify differentially rotating neutron stars with maximum mass among all types of solutions, which was the main goal of this article. Fig.~\ref{convD2} illustrates the precision and the method in the $(\rrat,\Hm)$ plane for type A stars with $\Atil=0.7$. Once a rough estimation of the position in the $(\rrat,\Hm)$ plane was obtained (after building sequences as described in the main text of the article), the code was used to find the value of $M_0$ for each configuration associated to $({\rrat\,}_i,{\Hm\,}_j)$, where $i=0,1..,i_{\rm max}$ and $j=0,1..,j_{\rm max}$, with typical values of $i_{\rm max}$ and $j_{\rm max}$ in the range $5-15$. Then the maximum mass was determined as the extremum of the $M_0(\rrat,\Hm)$ function in that region of the plane, the corresponding configuration being also identify. Notice that when looking for the extremum of a smooth function such as $M_0(\rrat,\Hm)$, a spectral algorithm could also be used. The accuracy reached for the $n$th spectral approximant of $M_{\rm 0max}$ is shown on the right panel of Fig.~\ref{convD2}, while the left panel presents the corresponding $\rrat$ and $\Hm$. All values were obtained for $i_{\rm max}=j_{\rm max}=10$.

\end{document}